\documentclass[twocolumn,english,prl]{revtex4}
\usepackage[T1]{fontenc}
\usepackage[latin9]{inputenc}
\usepackage{array}
\usepackage{amsmath}
\usepackage{graphicx}
\usepackage{amssymb}

\makeatletter

\providecommand{\tabularnewline}{\\}

\@ifundefined{textcolor}{}
{%
 \definecolor{BLACK}{gray}{0}
 \definecolor{WHITE}{gray}{1}
 \definecolor{RED}{rgb}{1,0,0}
 \definecolor{GREEN}{rgb}{0,1,0}
 \definecolor{BLUE}{rgb}{0,0,1}
 \definecolor{CYAN}{cmyk}{1,0,0,0}
 \definecolor{MAGENTA}{cmyk}{0,1,0,0}
 \definecolor{YELLOW}{cmyk}{0,0,1,0}
 }

\usepackage{babel}
\makeatother

\makeatother

\usepackage{babel}

\begin{document}

\title{Kondo Lattice Scenario in Disordered Semiconductor Heterostructures}

\author{Kusum Dhochak and V. Tripathi}

\affiliation{Department of Theoretical Physics, Tata Institute of Fundamental
Research, Homi Bhabha Road, Navy Nagar, Mumbai 400005, India}
\begin{abstract}
We study nuclear relaxation in the presence of localized electrons
in a two-dimensional electron gas in a disordered delta-doped semiconductor
heterostructure and show that this method can reliably probe their
magnetic interactions and possible long-range order. In contrast,
we argue that transport measurements, the commonly-employed tool,
may not sometimes distinguish between spatial disorder and long-range
order. We illustrate the utility of using the nuclear relaxation method
to detect long-range order by analysing a recent proposal made on
the basis of transport measurements, on the spontaneous formation
of a two-dimensional Kondo lattice in a 2D electron gas in a heterostructure. 
\end{abstract}
\maketitle
\emph{Introduction-} The possibility of long-range charge or magnetic
order of strongly-correlated electrons in mesoscopic devices, such
as Wigner crystals \cite{bello,grill}, charge density waves \cite{willett},
and Kondo lattices \cite{siegert}, has attracted a great deal of
attention in recent times. Theoretical interest in these systems stems
from the low-dimensionality which enhances quantum effects, while
the practical motivation comes from the tunability of the material
parameters by electrical means, which is not achievable in bulk materials.
Experimental probes for long-range order in mesoscopic devices have
been usually based on transport measurements \cite{siegert} as their
small size makes it difficult to employ standard bulk methods such
as diffraction and nuclear magnetic resonance (NMR)\cite{alloul-kondo,boyce-kondo,ross-CDW}.
However suitably adapted NMR methods are now beginning to emerge as
very promising tools for studying electron interactions in mesoscopic
systems -- recent work shows that nuclear polarization may be generated
\cite{yusa,wald,dixon,machida,tripathi-rashba} locally in such devices
and its relaxation can be feasibly detected \cite{cooper,nesteroff}
through two-terminal conductance measurements, and the behavior of
the nuclear relaxation rate conveys useful information about the electronic
state in the device. For example in the context of the decade-old
puzzle of the $0.7$ conductance anomaly in quantum point-contact
devices \cite{thomas}, NMR can be used to distinguish between three
incompatible contesting theories - a Kondo effect, a spin-incoherent
Luttinger liquid state, or a polarized electron liquid \cite{cooper}
even though transport properties are similar in the three scenarios.

In this paper, we study nuclear relaxation as a probe for long-range
magnetic (and crystalline) order of localized spins in a disordered,
metallic two-dimensional electron gas (2DEG) in a delta-doped heterostructure.
We find that the temperature dependence of the relaxation rate for
a disordered few-impurity system approximately follows a linear$-T$
law, while for strong enough inter-spin interactions, nuclear relaxation
in a regular array, or a Kondo lattice, will show an exponential increase,
$e^{A/T},$ with decreasing temperature. In contrast, we argue that
transport measurements will show no significant difference between
the two situations. As an application of our analysis, we discuss
a recent experimental claim \cite{siegert} based on transport measurements
in disordered GaAs/AlGaAs delta-doped heterostructures, on the spontaneous
formation of a Kondo lattice in the 2D electron gas in the heterostructure.

\emph{Experimental context-} Kondo lattice materials, such as heavy
fermion metals, are being intensely studied \cite{gegenwart,schroeder}
to understand the nature of the competition of the magnetic ordering
tendency of the localized electrons and the screening tendency (Kondo
effect) of the conduction electrons, close to quantum criticality.
A 2D Kondo lattice, if engineered in a heterotructure, would offer
the twin advantages of reduced dimensionality and tunability of parameters\cite{siegert},
and, as we show below, nuclear relaxation can be used to study these
systems.

In Ref. \cite{siegert} it was observed that the 2DEG conductance
showed an alternating splitting and merging of a zero bias anomaly
(ZBA) upon varying the gate voltage $V_{g}.$ The authors interpreted
these observations as evidence for the formation of a spin-1/2 Kondo
lattice embedded in a 2DEG with the following physical picture. Varying
the gate voltage affects the 2DEG density, which, in turn, controls
the sign of the RKKY exchange interaction, $J_{RKKY}(R_{ij})\sim(J^{2}\rho/R_{ij}^{2})\cos(2k_{F}R_{ij}),$
of the localized spins. Here $J$ is the Kondo coupling of the localized
spins with the conduction electrons, and $\rho$ is the density of
states at the Fermi energy.

Nevertheless, the observation of the ZBA splitting is not sufficient
to prove the existence of a Kondo lattice. Such an effect has been
observed in the context of double quantum dot (DQD) systems \cite{jeong,lopez-aguado},
and attributed to the competition of Kondo and interdot exchange interactions.
Even in a sample with a small number of localized spins, the Kondo
and RKKY competition will be dominated by the pairs of spins with
the strongest exchange interactions. 
To this end, we need to show that the nuclear relaxation rates have
qualitatively different signatures for the Kondo lattice and few Kondo
impurities scenarios.

\begin{figure}
\includegraphics[width=7cm]{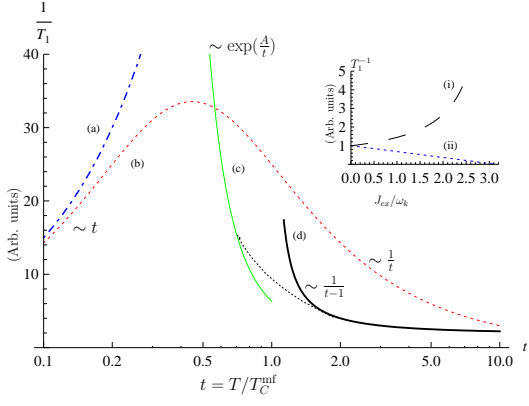}

\caption{(Color online) \label{fig:plots} Plots showing the qualitative differences
in the temperature and (AFM) interdot exchange interaction $J_{\text{ex}}$
dependencies of the nuclear relaxation rates $T_{1}^{-1}$ for a double
quantum dot system and Kondo lattice. Main plot: $T_{1}^{-1}(T)$
for (a) a double dot system; (b) a Kondo interaction dominated lattice
($J_{\text{ex}}/\omega_{K}<1$); (c,d) a Kondo lattice where $J_{\text{ex}}/\omega_{K}>1$
and $T<(>)T_{C}^{\text{mf}},$ where $T_{C}^{\text{mf}}$ is the mean-field
transition temperature. Dotted curve interpolates between these two
temperature regimes (there is no phase transition). Inset: $T_{1}^{-1}$as
a function of $J_{\text{ex}}/\omega_{K}$ for (i) the double dot system
- note that $T_{1}^{-1}$ vanishes for $J_{\text{ex}}/\omega_{K}>\pi;$
and (ii) for the Kondo lattice.}

\end{figure}

Nuclear relaxation takes place through nuclear coupling to localized
spins $\mathbf{S}$ as well as conduction electrons $\pmb\sigma:$
${\cal H}_{\text{loc}}=A_{d}\mathbf{I}\cdot\mathbf{S}+A_{s}\mathbf{I}\cdot\pmb\sigma.$
The relaxation contribution from localized (electron) spins is usually
much larger in devices similar to those considered here \cite{tripathi-kondo}.
Taking only the localized spin part, the nuclear relaxation rate can
be expressed in terms of the transverse impurity susceptibility, $T_{1}^{-1}=\frac{A_{d}^{2}k_{B}T}{\hbar^{2}(g_{s}\mu_{B})^{2}}\text{Im}\left(\frac{\chi_{i}^{+-}(\omega)}{2\omega}\right)_{\omega\rightarrow0}.$
We show below that the temperature dependencies of $1/T_{1}$ in the
Kondo lattice and few quantum dot scenarios are qualitatively different. The results
obtained in the paper are illustrated in Fig. \ref{fig:plots} and
Table \ref{tab:table}. %
\begin{table}
\begin{tabular}{|>{\centering}p{0.5in}|>{\raggedright}p{1.3in}|>{\raggedright}p{1.25in}|}
\hline 
 & {\small FM }  & {\small AFM}\tabularnewline
\hline 
{\small Double impurity }  & {\small Linear-$T$ at low temp. and $1/T$ at high temp. }  & {\small Zero at low temp. and $1/T$ at high temp.}\tabularnewline
\hline 
{\small Lattice }  & {\small $T/(T-T_{c})^{3/2}$ at high temp. and $\exp(1/T)$ at low
temp. }  & {\small $T/(T-T_{c})$ at high temp. and $\exp(1/T)$ at low temp. }\tabularnewline
\hline
\end{tabular}

\caption{\label{tab:table}Distinguishing different physical scenarios with
NMR}

\end{table}

\emph{Model-} We consider the following model Hamiltonian for $S=1/2$
magnetic impurities ${\mathbf{S}}_{i}$ in a 2-dimensional electron
gas: \begin{equation}
H=\sum_{k}\xi_{k}c_{k\sigma}^{\dagger}c_{k\sigma}+J\sum_{i}\pmb\sigma_{i}\cdot\mathbf{S}_{i}.\label{eq:hamiltonian}\end{equation}
 Here $\pmb\sigma_{i}$ the conduction electron spin density at $\mathbf{r}_{i}.$

We use the {}``drone-fermion'' representation for the localized
spins \cite{Mattis,Khaliullin}: $S_{i}^{+}=f_{i}^{\dagger}\chi_{i}/\sqrt{2},$
$S_{i}^{-}=\chi_{i}f_{i}/\sqrt{2},$ and $S_{i}^{z}=f_{i}^{\dagger}f_{i}-1/2,$
where $f_{i}$ and $f_{i}^{\dagger}$ are ferminonic operators and
$\chi$ are real Majorana fermions defined by $\{\chi_{i},\chi_{j}\}=\delta_{ij}$.
Note that the commutation relations for the impurity spins are automatically
satisfied, obviating the need to impose local constraints on the fermion
number. Introducing the bosonic operators, $a_{i}=(f_{i}^{\dagger}c_{i\uparrow}+\chi_{i}c_{i\downarrow}/\sqrt{2})/\sqrt{2},$
$b_{i}=(f_{i}^{\dagger}c_{i\downarrow}^{\dagger}-\chi_{i}c_{i\uparrow}^{\dagger}/\sqrt{2})/\sqrt{2},$
the interaction part of the Hamiltonian (up to constants and irrelevant
terms) can be written as $H_{int}=-J\sum_{i}\left(a_{i}^{\dagger}a_{i}+b_{i}^{\dagger}b_{i}\right).$
Factorizing $H_{int}$ using the Hubbard-Stratonovich transformation
introducing fields $\Delta_{1}^{i}$ and $\Delta_{2}^{i}$, and further
making the transformations $\Delta_{1,2}^{i}=|\Delta_{1,2}^{i}|e^{i\phi_{1,2}^{i}},$
$f\rightarrow fe^{i(\phi_{2}-\phi_{1})},$ $c_{\uparrow}(r)\rightarrow c_{\uparrow}(r)e^{i\phi_{2}},$
$c_{\downarrow}(r)\rightarrow c_{\downarrow}(r)e^{i\phi_{1}},$ the
partition function can be written in path integral form: \begin{align}
{\cal Z} & =\int D(c,f,\chi,\Delta)\, e^{-\int_{0}^{\frac{1}{T}}d\tau\left[S_{0}+S_{int}+\frac{1}{J}\sum_{i}\left(|\Delta_{1}^{i}|^{2}+|\Delta_{2}^{i}|^{2}\right)\right]};\nonumber \\
S_{0} & =\sum_{k\sigma}c_{k\sigma}^{\dagger}\left(\partial_{\tau}+\xi_{k}\right)c_{k\sigma}+\sum_{i}(f_{i}^{\dagger}\partial_{\tau}f_{i}+\frac{1}{2}\chi_{i}\partial_{\tau}\chi_{i}),\nonumber \\
S_{int} & =\sum_{i}|\Delta_{1}^{i}|\:(a_{i}+a_{i}^{\dagger})+|\Delta_{2}^{i}|\: (b_{i}+b_{i}^{\dagger})+\dot{\phi}\: \mbox{terms}.\label{eq:part-fn}\end{align}
 We make a mean field analysis, neglecting the fluctuations in $\phi$'s
and $\Delta$'s. The frequency-dependent local transverse susceptibility
at low temperatures $T\ll T_{K}$ can be shown to be \cite{Khaliullin}\begin{eqnarray}
\frac{\chi_{i}^{+-}\left(\omega_{m}\right)}{(g_{s}\mu_{B})^{2}}=\langle T_{\tau}S_{i}^{+}(\tau)S_{i}^{-}\left(\tau\prime\right)\rangle_{\omega_{m}} & \simeq & \frac{2}{\pi\left(|\omega_{m}|+\omega_{K}\right)}.\label{eq:imp-susc-lowT}\end{eqnarray}
 Here $\omega_{m}$ are bosonic Matsubara frequencies, and $\omega_{K}=D\exp(-4/3\rho J).$
Note that $\omega_{K}$ differs from the correct Kondo temperature,
$k_{B}T_{K}\sim De^{-1/(\rho J)}.$ This is an artifact of the mean field approach. An
analytic continuation, $\chi_{i}^{+-}(\omega_{m})\rightarrow\chi_{i}^{+-}(\omega)=(g_{s}\mu_{B})^{2}/\pi(-i\hbar\omega+\omega_{Ki}),$
to real frequencies leads to the well-known result $T_{1i}^{-1}=A_{d}^{2}k_{B}T/\pi\hbar\omega_{Ki}^{2}$
for $T\ll\omega_{K}.$

Nuclei may also relax through their hyperfine coupling with conduction
electrons. It is easy to see that at low temperatures, the ratio of
the nuclear relaxation rates from impurity coupling and conduction
electron coupling is $(\omega_{K}^{2}\rho^{2}R_{en}^{4}\pi)^{-1},$
where $R_{en}$ is the electron-nucleus separation. Thus the impurity
coupling mechanism dominates as long as $R_{en}<1/\sqrt{\omega_{K}\rho\pi^{1/2}}\approx160{\rm nm}.$

\emph{Double or a few-impurity system-} We now consider the impurity
spin susceptibility for two spins $\mathbf{S}_{1},\mathbf{S}_{2}$
at $\mathbf{R}=\mathbf{R}_{1},\mathbf{R}_{2}$ which have an exchange
interaction $H_{\text{ex}}=J_{\text{ex}}(R_{12})\mathbf{S}_{1}\cdot\mathbf{S}_{2}$
among them. If the wavefunctions of the localized electrons have a
significant overlap, then direct exchange would be dominant. Indirect
(or RKKY) exchange is more important at larger separations. Here it
also becomes important to compare the relative strengths of the RKKY
interaction between the impurity spins with the hyperfine interaction
of either of the impurities with neighboring nuclei. The RKKY interaction
$J_{\text{RKKY}}$ falls off with distance $R_{12}$ not faster than
$J_{\text{RKKY}}\sim J^{2}\rho/R_{12}^{2}.$ This should be compared
with $A_{d}=A_{s}/l_{\text{loc}}R_{\text{dot}}^{2},$ where $R_{\text{dot}}$
is the size of the quantum dot in the plane of the heterostructure
and $l_{\text{loc}}$ is the thickness of the 2DEG. We use the following
parameters for a GaAs/AlGaAs heterostructure, $J\rho\sim1,$ $l_{\text{loc}}\sim1{\rm nm},$
$R_{\text{dot}}\sim10{\rm nm},$ $A_{s}=3.8\times10^{-54}{\rm Jm}^{3},$
and $m=0.063m_{e}.$ Then $J_{\text{RKKY}}\gg A_{d}/N_{\text{nuc}}$
is satisfied if $R_{\text{12}}\ll\sqrt{1/A_{d}\rho}\approx1{\rm mm},$
and this is true for most devices. Thus the nuclei couple to the RKKY
bound pair $\mathbf{S}_{1}+\mathbf{S}_{2}$ rather than the spins
separately. The impurity susceptibility now involves both on-site and intersite 
correlations, and we have, to leading order in inter-impurity interaction
for $T\ll T_{K}$, \begin{equation}
\begin{split}T_{1}^{-1}=\frac{A_{d}^{2}k_{B}T}{\pi\hbar\omega_{K1}^{2}}\left(1+\frac{\omega_{K1}^{2}}{\omega_{K2}^{2}}-\frac{J_{\text{ex}}}{\pi\omega_{K2}}\left(1+\frac{\omega_{K1}}{\omega_{K2}}\right)\right)\end{split}
.\label{eq:T1-two-imp-dissimilar}\end{equation}

When $J_{\text{ex}}/\omega_{K}\geq\pi,$ the nuclear relaxation rate
is suppressed to zero: this is the maximum value of $J_{\text{ex}}/\omega_{K}$
for which the behavior is governed by the Kondo screening of the impurity
spins. Indeed, even for a large \emph{ferromagnetic} coupling of the
spins, the ground state is a Kondo singlet \cite{jones}. At antiferromagnetic
couplings $J_{\text{ex}}>\pi\omega_{K},$ the ground state is an RKKY
singlet which is unable to exchange spins with the nuclei. While our
analysis is only to leading order in $J_{\text{ex}},$ more accurate
calculations \cite{jones} based on numerical renormalization group
methods have shown that this critical point occurs at $J_{\text{ex}}/k_{B}T_{K}\approx2.2.$

We discuss now the validity of the mean field treatment. Note that
the mean field $\Delta$ corresponds to binding energy of the impurity
fermion with the local conduction electron. Ignoring fluctuations
of the phase of $\Delta$ results in underestimation of $\omega_{K}$
(see text following Eq. \ref{eq:imp-susc-lowT} and also Ref. \cite{Khaliullin}).
Amplitude fluctuations of $\Delta$ may be ignored as long as the
Kondo energy dominates inter-impurity exchange, i.e., for small $J_{\text{ex}}/\omega_{K}.$
In fact, the mean field approach is incapable of capturing the physics
of the magnetically ordered phase.

With a larger number of spatially disordered impurity spins, one can
show that for weak inter-impurity interactions, the nuclear relaxation
rates have linear-$T$ behavior with logarithmic factors arising from
the random distribution of Kondo temperatures of individual impurities
\cite{miranda}. For strong inter-impurity exchange interactions,
we can ignore the Kondo effect to leading order. In that case, it
is known that the magnetic susceptibility at low temperatures is dominated
by pairs with the weakest exchange interactions \cite{bhatt-fisher}
-- this leads to a weakly-increasing susceptibility $e^{C\ln^{1/2}(T_{0}/T)}$
(instead of zero for the double impurity case). Nevertheless, the
nuclear relaxation rate is dominated by the linear-$T$ prefactor
as the exponential term is weaker than any power law.

\emph{Kondo lattice-} The main physical difference from the two-impurity
case is the existence of low energy magnetic excitations in the lattice
for any value of the ratio $J_{\text{ex}}/\omega_{K}.$ As a result,
significant nuclear relaxation still occurs for large antiferromagnetic
inter-impurity couplings unlike the two-impurity case where it vanishes.

We consider first the scenario where we have a lattice of Kondo impurities
with a weak exchange interaction ($J_{\text{ex}}\ll\omega_{K}$) among
the neighboring spins. Suppose that $J_{\text{ex}}(\mathbf{q})$ has
maximum value at $\mathbf{q}=\mathbf{Q},$ and assume the wavevector
dependence in the vicinity of the maximum is $J_{\text{ex}}(\mathbf{Q}+\mathbf{q})=J_{\text{ex}}(\mathbf{Q})-(D_{s}/n_{\text{imp}})a^{2}q^{2},$
where $a$ is the lattice constant of the Kondo array and $D_{s}$
the spin wave stiffness. The random phase approximation (RPA) susceptibility
in this momentum region has the form \begin{eqnarray}
\chi_{\mathbf{Q}+\mathbf{q}}^{+-}(\omega,T) & = & \frac{(g_{s}\mu_{B})^{2}}{\pi\left(\omega_{sf}\left(T\right)-i\hbar\omega+\frac{D_{s}a^{2}q^{2}}{\pi}\right)},\label{eq:chi-sf}\end{eqnarray}
 where $\omega_{sf}\left(T\right)=\frac{(g_{s}\mu_{B})^{2}}{\pi\chi_{i}^{+-}(0,T)}-\frac{J_{\text{ex}}(Q)n_{\text{imp}}}{\pi}.$
A new energy scale $\omega_{sf}(0)=\omega_{K}-J_{\text{ex}}(Q)n_{\text{imp}}/\pi$
appears representing the competition of Kondo and inter-impurity exchange
interactions. As $J_{\text{ex}}(Q)n_{\text{imp}}\rightarrow\pi\omega_{K},$
the uniform, static transverse susceptibility tends to diverge signaling
a magnetic phase transition. Using the known temperature dependence
of the susceptibility of a Kondo impurity, $\chi_{i}^{+-}(0,T)\simeq\chi_{i}(0)(1-Ck_{B}^{2}T^{2}/\omega_{K}^{2}),$($C$
is a constant of order $1$), together with the frequency dependence
of $\text{Im}\chi$ from Eq. \ref{eq:chi-sf}, the nuclear relaxation
rate for $k_{B}T\ll\omega_{K}$ turns out to be \begin{equation}
T_{1}^{-1}=A_{d}^{2}k_{B}T/4\pi^{2}\hbar D_{s}\omega_{sf}(T).\label{eq:T1-weak-J}\end{equation}
 There is a crucial difference between the nuclear relaxation results
for the Kondo lattice in Eq. \ref{eq:T1-weak-J} and the two-impurity
case. Consider for simplicity $\omega_{K1}=\omega_{K2}=\omega_{K}.$
First, near the transition $J_{\text{ex}}(Q)n_{\text{imp}}/\omega_{K}=\pi,$
$1/T_{1}$ for the Kondo lattice is large and finite, while it tends
to vanish for the two-impurity case.

Now we consider the case when localized spin-spin interaction is dominant
and neglect the Kondo interaction in the zeroth order. We are particularly
interested in the regime close to a magnetic phase transition. The
Hamiltonian describing the system would be $H=\sum_{k,\sigma}\xi_{k}c_{k\sigma}^{\dagger}c_{k\sigma}+\sum_{\mathbf{q}}J_{\text{ex}}(\mathbf{q})\mathbf{S}_{\mathbf{q}}\cdot\mathbf{S}_{-\mathbf{q}}+J\sum_{i}\pmb\sigma_{i}\cdot\mathbf{S}_{i},$
where $J$ is to be treated now as a perturbation. $J_{\text{ex}}(\mathbf{q})$
represents all exchange processes except indirect exchange (RKKY),
$J_{\text{RKKY}}(\mathbf{q},\omega)=J^{2}\sum_{\mathbf{k}}\frac{n_{\mathbf{k}-\mathbf{q}/2}-n_{\mathbf{k}+\mathbf{q}/2}}{\omega+\epsilon_{\mathbf{k}-\mathbf{q}/2}-\epsilon_{\mathbf{k}+\mathbf{q}/2}+i\delta}.$
Thus we may write the effective inter-impurity exchange interaction
as $J_{\text{ex}}(q,\omega)=J_{\text{ex}}(q)+J_{\text{RKKY}}(q,\omega).$ where 
We now Taylor expand the exchange interaction near its extremum, $J_{\text{ex}}(\mathbf{Q}+\mathbf{q})=J_{\text{ex}}(\mathbf{Q})(1-\alpha^{2}q^{2}),$ where $J_{\text{ex}}(\mathbf{Q})\alpha^{2}=(D_{s}/n_{\text{imp}})a^{2}.$
The spin susceptibility near the ordering point is approximately \begin{eqnarray}
\chi_{\mathbf{Q}+\mathbf{q}}^{+-}(\omega) & = & \frac{(g_{s}\mu_{B})^{2}}{4k_{B}T_{C}^{\text{mf}}\left\lbrace \alpha^{2}/\xi^{2}+\alpha^{2}q^{2}-i\gamma_{\mathbf{Q}+\mathbf{q}}(\omega)\right\rbrace },\label{eq:chi-Q-large-J}\end{eqnarray}
 where $\mathbf{Q}$ is the wave vector of ordering, $T_{C}^{\text{mf}}$
is the mean-field magnetic transition temperature, $\xi(T)$ is the
magnetic correlation length and $\gamma_{\mathbf{q}}(\omega)$ is
the imaginary part of $J_{\text{ex}}(\mathbf{q},\omega)/J_{\text{ex}}(\mathbf{Q}),$
$\gamma_{\mathbf{q}}(\omega)\simeq\pi(J\rho)^{2}\hbar\omega/4J_{\text{ex}}(\mathbf{Q})k_{F}q=\gamma(q)\omega.$
We can now estimate the nuclear relaxation rate. For an antiferromagnetic
(AFM) square lattice, the ordering happens at $\mathbf{Q}=(\pi/a,\pi/a).$
Then the relaxation rate at the site of any given impurity is \begin{align}
T_{1}^{-1}(T) & =\frac{A_{d}^{2}(J\rho)^{2}J_{\text{ex}}(Q)n_{\text{imp}}^{2}}{64\hbar D_{s}^{2}k_{F}Q}\frac{T}{T_{C}^{\text{mf}}}\:\frac{\xi(T)^{2}}{a^{2}}.\label{eq:T1-afm}\end{align}
 Eq. \ref{eq:T1-afm} differs from estimates \cite{chakravarty} of
$1/T_{1}$ for Heisenberg antiferromagnets because in our case, the
magnon decay is on account of the RKKY coupling of the impurity spins.
Similarly for the ferromagnetic (FM) case, \begin{eqnarray}
T_{1}^{-1}(T) & \approx & \frac{A_{d}^{2}\pi(J\rho)^{2}J_{\text{ex}}(Q)n_{\text{imp}}}{128\hbar k_{F}aD_{s}^{2}}\frac{T}{T_{C}^{\text{mf}}}\frac{\xi(T)^{3}}{a^{3}}.\label{eq:T1-fm}\end{eqnarray}
 The temperature dependencies of the correlation lengths are similar
for the AFM and FM cases, \begin{align}
\xi(T) & \simeq\left\{ \begin{array}{l}
\alpha\sqrt{T_{C}^{\text{mf}}/(T-T_{C}^{\text{mf}})},\qquad T>T_{C}^{\text{mf}},\\
a\exp(2\pi D_{s}'/k_{B}T),\qquad T<T_{C}^{\text{mf}},\end{array}\right.\label{eq:correl-fm}\end{align}
 where the low temperature behavior for the antiferromagnet was obtained
in Refs. \cite{chakravarty,arovas}, and for the ferromagnet
from Refs. \cite{arovas,takahashi}. $D_{s}'\approx0.18J_{\text{ex}}(\mathbf{Q})n_{\text{imp}}$
is the exact spin wave stiffness at $T=0$ for a 2D (square lattice)
Heisenberg magnet. These results also differ from the usually-encountered
3D Kondo lattice systems \cite{moriya}, because of the qualitative
difference in the behavior of $\xi(T)$ at low temperatures.

Finally let us discuss the effect of the Kondo interaction on our
results. In presence of inter-impurity exchange interactions, the
singular Kondo corrections ($\sim(J\rho)\ln(D/k_{B}T)$) to the gyromagnetic
ratio of the impurity spins are modified to $(J\rho)\ln(D/\sqrt{J_{\text{ex}}^{2}+k_{B}^{2}T^{2}})$ \cite{tsay}.
Consequently, the primary effect of Kondo corrections is to decrease
the Stoner critical temperature $T_{C}^{\text{mf}}$ as well as the
pre-factor in the expressions for the nuclear relaxation rates but
the temperature dependence of $T_{1}^{-1}$ does not change significantly.
Eqs. \ref{eq:T1-two-imp-dissimilar}, \ref{eq:T1-weak-J}, \ref{eq:T1-afm}
and \ref{eq:T1-fm} are our main results and are plotted in Fig. \ref{fig:plots}.

There is also a possibility of a magnetic instability below the Kondo
temperature. In this case the ZBA splitting would be smaller than
the {}``heavy fermion'' bandwidth, $\omega_{K}$, unlike the above
case where the ZBA splitting is larger. As an example, for a Fermi
liquid with FM spin fluctuations, results for nuclear relaxation available
in the literature \cite{hatatani} and are similar to our $J_{RKKY}/\omega_{K}>1$
results. The difference will be seen in the magnitude of ZBA splitting
relative to $\omega_{K}$.

In summary, we calculated the nuclear relaxation rates $T_{1}^{-1}$
for the Kondo lattice and the few disordered magnetic impurities cases
and showed that they have qualitatively different low temperature
behaviors: when inter-spin exchange interactions are strong compared
to the Kondo energy $\omega_{K},$ the temperature dependence of $T_{1}^{-1}$
for the few-impurity system will follow an approximate linear$-T$
law, while for the Kondo lattice $T_{1}^{-1}$ will show an exponential
behviour $e^{A/T}$ at low temperatures. In contrast, we
argued that transport measurements \cite{siegert} in this case may
not provide a clinching evidence for the formation of crystalline
order (Kondo lattice). The exponential temperature dependence is special
to two dimensions and indicates stronger spin fluctuations: a power-law
behavior is expected in three dimensions on either side of the transition
temperature\cite{moriya}. These results also differ from a 2D Heisenberg
magnet because in our case, magnon decay is mediated by conduction
electrons. We hope our study will work towards encouraging the use
of NMR measurements as an additional handle for studying magnetism
and long-range order in low-dimensional conductors.

The authors benefited from discussions with M. Kennett and N. Cooper.
K.D. and V.T. thank TIFR for support. V.T. also thanks DST for a Ramanujan
Grant {[}No. SR/S2/RJN-23/2006{]}.

\end{document}